\documentclass[slac_one]{revtex4}
\usepackage{graphicx}
\usepackage{fancyhdr}
\begin{document}

\title{Observation of the bottomonium ground state, \boldmath $\eta_b$, \unboldmath at BaBar}

\author{P. Grenier}
\affiliation{SLAC, 2575 Sand Hill Road, Menlo Park, CA 94025, USA}
\affiliation{Representing the BaBar Collaboration}

\begin{abstract}
We present the first observation of the bottomonium ground state $\eta_b(1S)$ 
in the photon energy spectrum using a sample of $(109 \pm 1)$ 
million of $\Upsilon(3S)$ events recorded at the $\Upsilon(3S)$ energy with the BaBar detector 
at the PEP-II $B$ factory at SLAC. 
A peak at $E_\gamma = 921.2 ^{+2.1}_{-2.8} {\rm (stat)}\pm 2.4{\rm (syst)}$ MeV observed 
with a significance of 10 standard deviations in the photon energy spectrum is interpretated 
as being due to the radiative transition 
$\Upsilon(3S) \to \gamma \, \eta_b(1S)$. This photon energy corresponds to an $\eta_b(1S)$ mass of 
$9388.9 ^{+3.1}_{-2.3} {\rm (stat)} \pm 2.7{\rm (syst)}$ MeV/$c^2$. The hyperfine 
$\Upsilon(1S)$-$\eta_b(1S)$ mass splitting is 
$71.4 ^{+2.3}_{-3.1} {\rm (stat)} \pm 2.7{\rm (syst)}$ MeV/$c^2$. The branching fraction 
for this radiative $\Upsilon(3S)$ decay is obtained as 
$(4.8 \pm 0.5{\rm (stat)}  \pm 1.2 {\rm (syst)}) \times 10^{-4}$.
\end{abstract}

\maketitle

\thispagestyle{fancy}

\section{INTRODUCTION}

Bottomonium spectroscopy started thirty years ago with the discovery of the $\Upsilon(nS)$ 
resonances~\cite{ref:Ydiscovery}. The spin-singlet states $h_b(nP)$ and $\eta_b(nS)$ have 
yet to be observed. In particular, the ground state of the bottomonium spectrum, $\eta_b(1S)$, 
was still missing. 
The mass difference between the $\Upsilon(1S)$ and the $\eta_b(1S)$, the hyperfine splitting, 
is very important in understanding the role of spin-spin interaction in heavy quark bound 
systems and in testing calculations and predictions from various models such as Quark Models, 
pNRCQCD and Lattice QCD~\cite{QWG-YR}. Predictions for the hyperfine splitting vary from 
36 to 100 MeV/$c^2$~\cite{ref:GodfreyRosner}.

We report on the observation of the bottomonium ground state $\eta_b$ from the radiative 
transition $\Upsilon(3S) \rightarrow \gamma \eta_b$ ~\cite{ref:ourPRL}. Theoritical predictions 
for the branching fraction of the decay vary from 1 to 20$\times 10^{-4}$\cite{ref:GodfreyRosner}. 
The CLEO III experiment has published a 90$\%$ confidence level upper limit for the branching 
fraction ${\cal B}[\Upsilon(3S)\to \gamma\, \eta_b] <4.3\times10^{-4}$~\cite{ref:cleo}.

The data used in this study was recorded with the BaBar detector~\cite{ref:babar} at the 
PEP-II asymmetric-energy $e^+e^-$ storage rings. It consists of 28.0 fb$^{-1}$ of integrated 
luminosity collected at a $e^+e^-$ CM energy of 10.355~GeV, corresponding to the mass of the 
$\Upsilon(3S)$ resonance. Samples of 2.4 fb$^{-1}$ and 43.9 fb$^{-1}$ recorded 
30~MeV below the $\Upsilon(3S)$ and 40~MeV below the $\Upsilon(4S)$ resonances were used for background 
studies. The trajectories of charged particles are reconstructed using a combination of five layers of 
double-sided silicon strip detectors and a 40-layer drift chamber, all operated inside the 1.5-T 
magnetic field of a superconducting solenoid. Photons are detected using a CsI(Tl) electromagnetic 
calorimeter (EMC), which is also inside the coil. The energy resolution for photons varies from 
2.9\% (at 600 MeV) to 2.5\% (at 1400 MeV).

\section{BACKGROUNDS AND SIGNAL SELECTION}

The signal for $\Upsilon(3S) \to \gamma \, \eta_b$ is extracted from a binned maximum likelihood 
fit to the inclusive photon energy spectrum in the center of mass (CM) frame. The monochromatic 
photon from the decay will appear as a bump in the photon energy ($E_\gamma$) distribution. For an 
$\eta_b$ mass of 9.4 GeV, and the $\Upsilon(3S)$ energy, the photon energy shall peak at 911 MeV. 
We are therefore looking for an enhancement in the $E_\gamma$ distribution near 900 MeV.

\subsection{Background contributions to the \boldmath $E_\gamma$ \unboldmath distribution}

There are two main background contributions to the photon energy distribution
The first contribution produces a smooth non-peaking background. It comes from continuum 
events ($e^+e^- \to q\bar q$ where $q=u,d,s,c$) and bottomonium decays. 
The second contribution produces peaks in the $E_\gamma$ spectrum, close to the expected 
signal position. It comes from two processes:

\begin{itemize}

\item 
The exclusive decay 
$\Upsilon(3S) \to \gamma \chi_{bJ}(2P); \chi_{bJ}(2P)\rightarrow\gamma\Upsilon(1S), \ J=0, 1, 2$. The 
second radiative transitions produce a broad peak centered at 760 MeV. As there are three transitions, we 
would expect to observe three peaks. However, due to the detector energy resolution and to the 
Doppler broadening, that arises from the motion of the $\chi_{bJ}(2P)$ states in the $\Upsilon(3S)$ CM 
frame, the three peaks merge into a single broad bump. 

\item
The radiative production of the $\Upsilon(1S)$ through initial state radiation (ISR): 
$e^+e^- \to \gamma_{ISR} \, \Upsilon(1S)$. This process produces a peak centered at 856 MeV. 

\end{itemize}

In order to extract the $\eta_b$ signal, it is crucial to understand both the 
lineshapes and the yields of the two peaking background components.

\subsection{Signal Selection}

The selection criteria have been optimized by maximazing the figure of merit $S/ \sqrt{B}$, where $S$ 
and $B$ represent the expected yield for signal and background respectively. The signal sample is 
obtained from a detailed Monte Carlo (MC) simulation. There is no reliable event generator to 
model the various bottomonium decays. A small fraction ($9 \%$) of the data sample 
(in the region $0.85<E_\gamma<0.95$ GeV) was used to model the background. In order to avoid any 
bias, this small data set was not used for the extraction of the signal in the final fit.

As the $\eta_b$ is expected to decay mainly through two gluons, one can expect a large 
track multiplicity in the final state. Events are selected by requiring at least four tracks 
in the event and that the ratio of the second to zeroth Fox-Wolfram moments~\cite{ref:fox} be 
less than 0.98. 

Photons are first required to be isolated from all charged tracks, and their shapes are required 
to be consistent with an electromagnetic shower: the lateral moments~\cite{ref:LAT} are required 
to be less than 0.55. In order to reduce the contribution from ISR events 
$e^+ e^- \rightarrow \gamma_{ISR} \Upsilon(1S)$, candidate photons are required to be detected 
in the central region of the calorimeter $-0.762<\cos(\theta_{\gamma, LAB})<0.890$, where 
$\theta_{\gamma, LAB}$ is the angle between the photon and the beam axis in the laboratory frame.

We apply a cut on the angle $\theta_T$ between the direction of the photon momemtum and the thrust 
axis~\cite{ref:brandt}. The thrust axis is computed with all charged tracks and neutral calorimeter 
clusters in the event, excluding the photon candidate. Given that the $\eta_b$ is a spin-zero 
resonance, the angle distribution for the signal should be flat. However, for continuum events the 
distribution should be peaking at the forward and backward directions. We require 
$|\cos{\theta_T}|< 0.7$.

Finally we apply a veto to reduce photons coming from $\pi^0$ decays. These photons represents the main 
source of background. A photon candidate combined with any photons in the event is required not to 
have an invariant mass within 15 MeV of the nominal $\pi^0$ mass. The energy of the second photon in 
the $\pi^0$ candidate  is required to be larger than 50 MeV. 

These selection criteria lead to an efficiency of $37 \%$ and $6\%$ on signal and background respectively.

The optimization procedure was checked on data using the broad peak from the second radiative 
transition of the $\Upsilon(3S) \to \gamma \chi_{bJ}(2P); \chi_{bJ}(2P)\rightarrow\gamma\Upsilon(1S)$ 
process. It yielded to a very similar cut optimization.

\section{FITTING PROCEDURE}

\subsection{Introduction}

The $\eta_b$ signal is extracted, after all selection cuts are applied,  using a 
binned maximum likelihood fit to the inclusive photon energy spectrum in the CM frame in the range 
$0.5<E_\gamma<1.1$ GeV. 

There are four components to the fit:

\begin{itemize}

\item non-peaking background;

\item $\chi_{bJ}(2P)\rightarrow \gamma \, \Upsilon(1S)$ peaking background;

\item $\gamma_{ISR} \Upsilon(1S)$ peaking background;

\item $\eta_b$ signal.

\end{itemize}

\begin{figure*}[t]
\centering
\includegraphics[width=95mm]{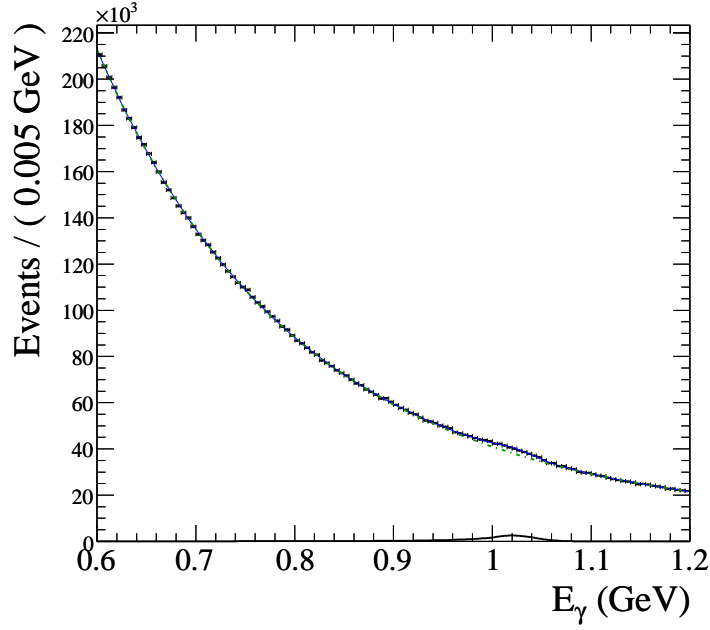}
\includegraphics[width=85mm]{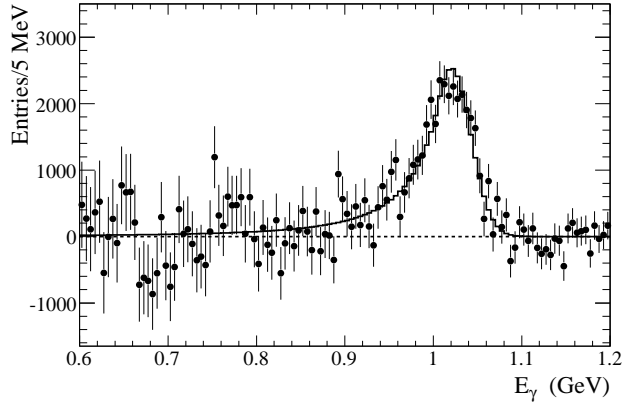}
\caption{The top figure shows the $E_\gamma$ distribution from the $\Upsilon(4S)$ Off-Peak 
data sample. The bottom figure shows the background subtracted distribution. The 
$\gamma_{ISR} \Upsilon(1S)$ peak is clearly visible.} \label{fig1_ISR}
\end{figure*}

\subsection{Probability density functions}

The non-peaking backround has been parametrized by the following probability 
density function: $f(E_{\gamma}) = A \left( C + {\rm exp}[-\alpha E_\gamma - \beta E^2_\gamma] \right) $.

\begin{figure*}[t]
\centering
\includegraphics[width=95mm]{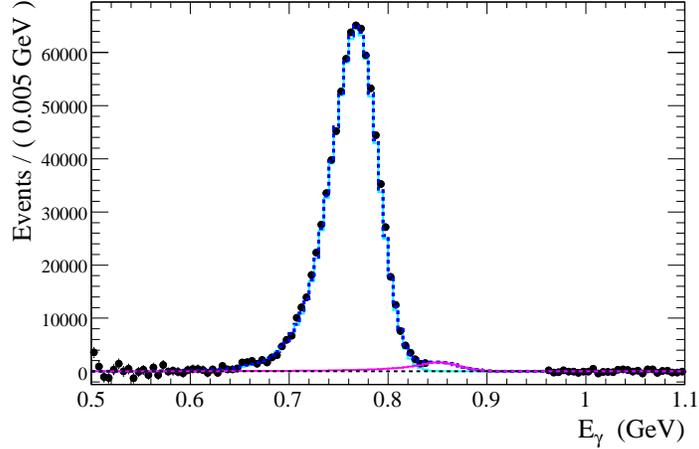}
\caption{Fit to the background subtracted $E_\gamma$ distribution, with the signal 
excluded, for the determination of the $\chi_{bJ}(2P)\rightarrow \gamma \, \Upsilon(1S)$ 
PDF parameters. The blue dotted line shows the $\chi_{bJ}(2P)\rightarrow \gamma \, \Upsilon(1S)$ 
lineshape and the full purple line shows the $\gamma_{ISR} \Upsilon(1S)$ lineshape.} \label{fig_chib}
\end{figure*}

As explained above, due to detector energy resolution and Doppler broadening, the three 
peaks from the $\chi_{bJ}(2P)\rightarrow \gamma \, \Upsilon(1S)$ transitions are merged. 
The three peaks have been modeled using a Gaussian modified with a power-law tail on the low side 
(Crystall Ball (CB) function~\cite{ref:CB}). The relatives rates and peak positions 
between the three peaks have been fixed from the PDG values~\cite{ref:PDG}. For each $\chi_{bJ}(2P)$ 
lineshape, the parameters of the 
power-law tail have been fixed to a commom value. The PDF parameters have been determined from 
fitting non-peaking background subtracted the $E_\gamma$ distribution where the signal region 
(840 to 960 MeV) has been excluded (see Figure \ref{fig_chib}). For the final fit, all the 
$\chi_{bJ}(2P)\rightarrow \gamma \, \Upsilon(1S)$ component parameters were fixed to the values 
obtained from this fit, 
except the yield. 

The PDF for the $\gamma_{ISR} \Upsilon(1S)$ (i.e. ISR peak) peaking background component 
was modeled with a CB function. All the CB parameters were obtained from MC. Given the detector energy resolution, 
the ISR and signal peaks are likely to overlap. Depending on the mass of the $\eta_b$, the overlap could 
be large. In the final fit, it was therefore decided to fix the yield of the ISR peak. The rate of the 
ISR peak was estimated using data taken 40 MeV below the $\Upsilon(4S)$ resonance ($\Upsilon(4S)$ Off-Peak 
data). The top plot of Figure \ref{fig1_ISR} shows the $E_\gamma$ distribution for the $\Upsilon(4S)$ 
Off-Peak data after all cuts are applied. The bottom plot shows the same distribution after subtracting 
the non-peaking background. A clear ISR peak is seen. A fit with a CB functions yields to 
$35800 \pm 1600$ events. The yield is extrapolated to the $\Upsilon(3S)$ energy using the relative 
cross-sections, integrated luminosities and signal reconstruction efficiencies. The estimated 
$\gamma_{ISR} \Upsilon(1S)$ yield is then $25200 \pm 1700$ events. The error includes
systematic uncertainties. This is consistent with but more precise than the yield estimated with 
data taken below the $\Upsilon(3S)$ resonance. 

The $\eta_b$ PDF is modeled with a non-relativistic Breit-Wigner funtion (for the natural shape 
of the $\eta_b$) convolved with a CB function which models the energy resolution. The CB paramaters 
were fixed from MC. MC experiments have shown that the width of the $\eta_b$ had to be fixed 
in the final fit. It is not known, but theoritical prediction vary from 4 to 20 MeV~\cite{ref:widththeory}. 
We used a value of 10 MeV for the nominal fit.

\subsection{Fit to the full data sample}

\begin{figure*}
\centering
\includegraphics[width=85mm]{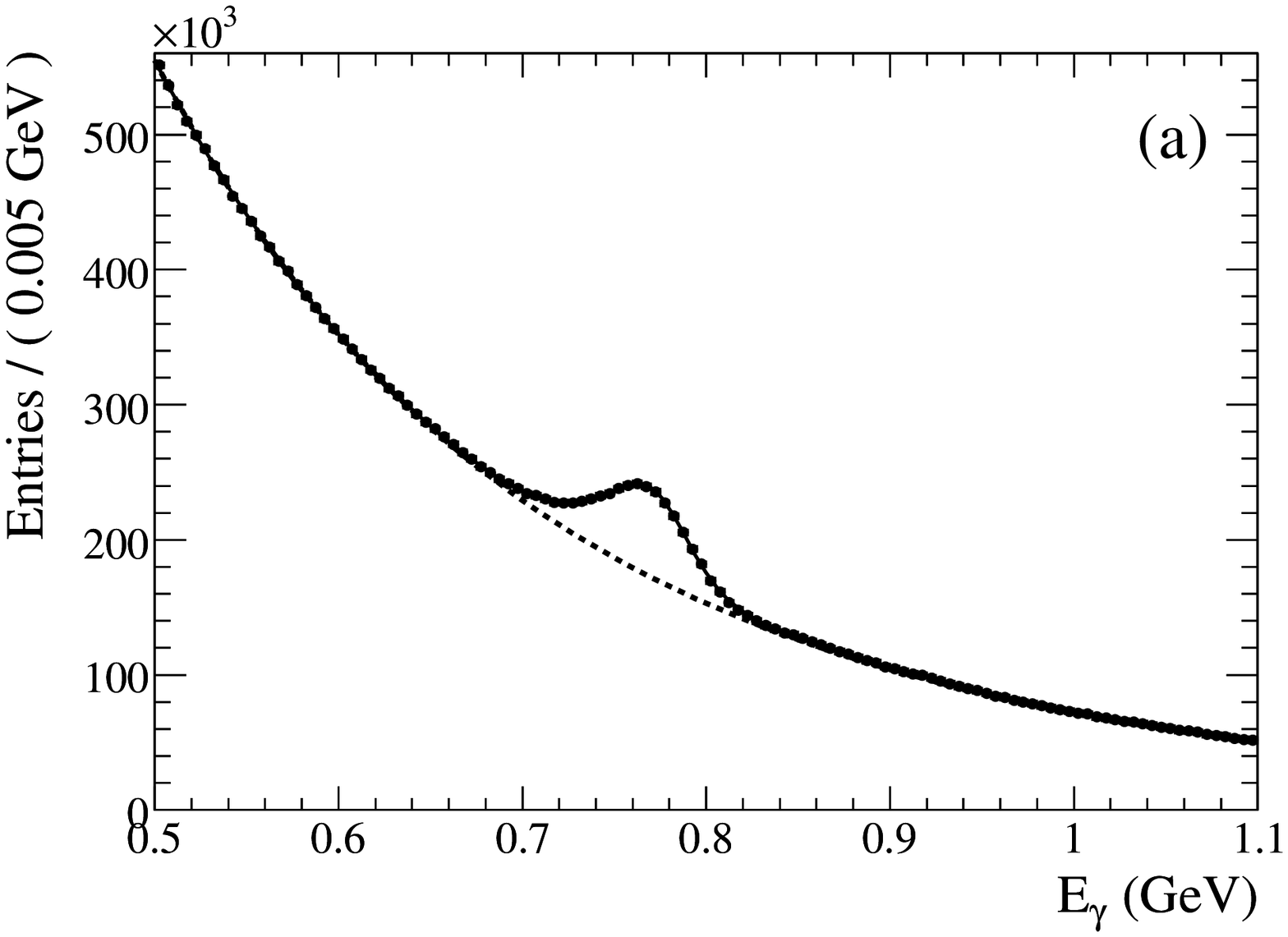}
\includegraphics[width=90mm]{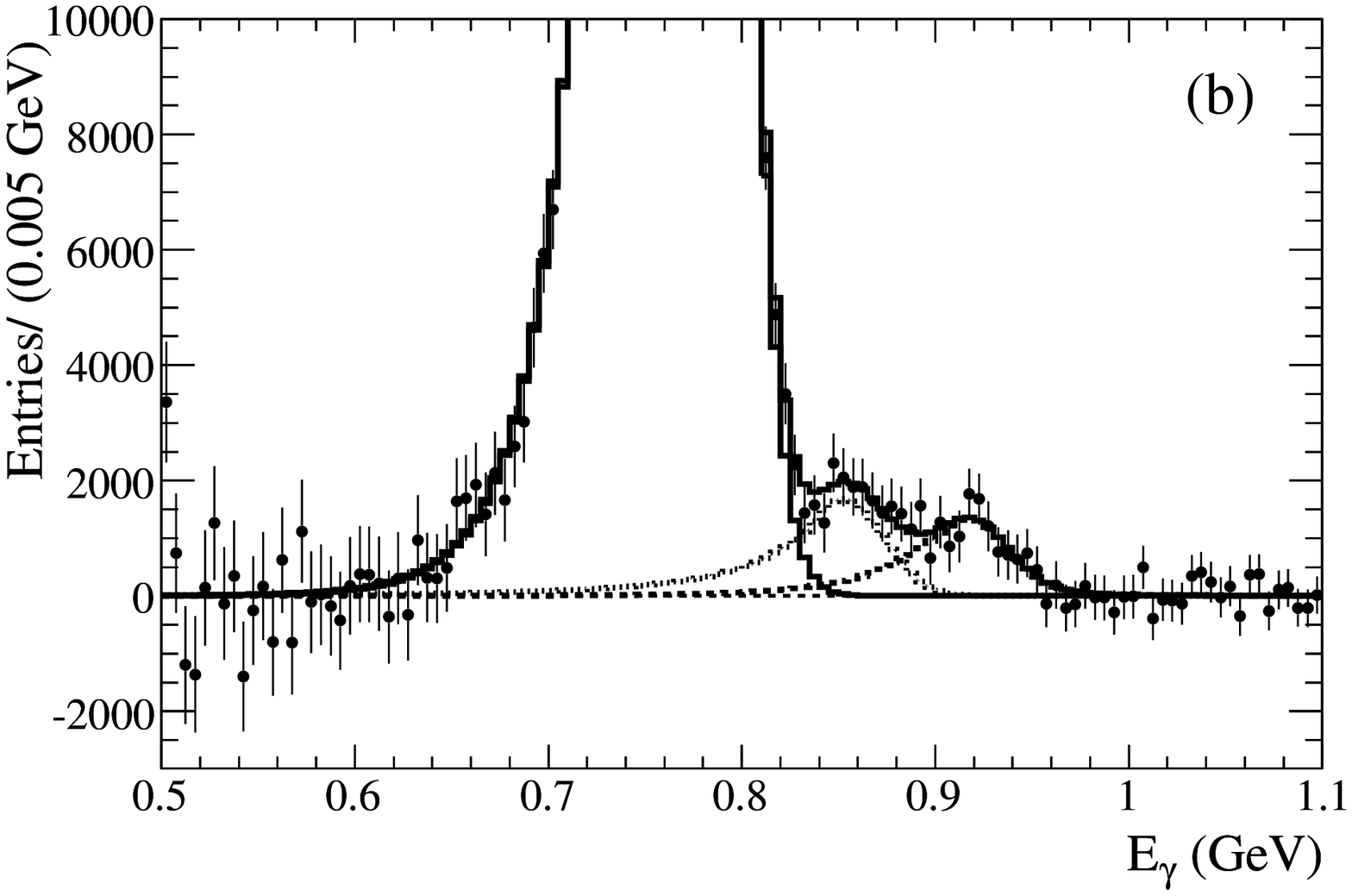}
\includegraphics[width=90mm]{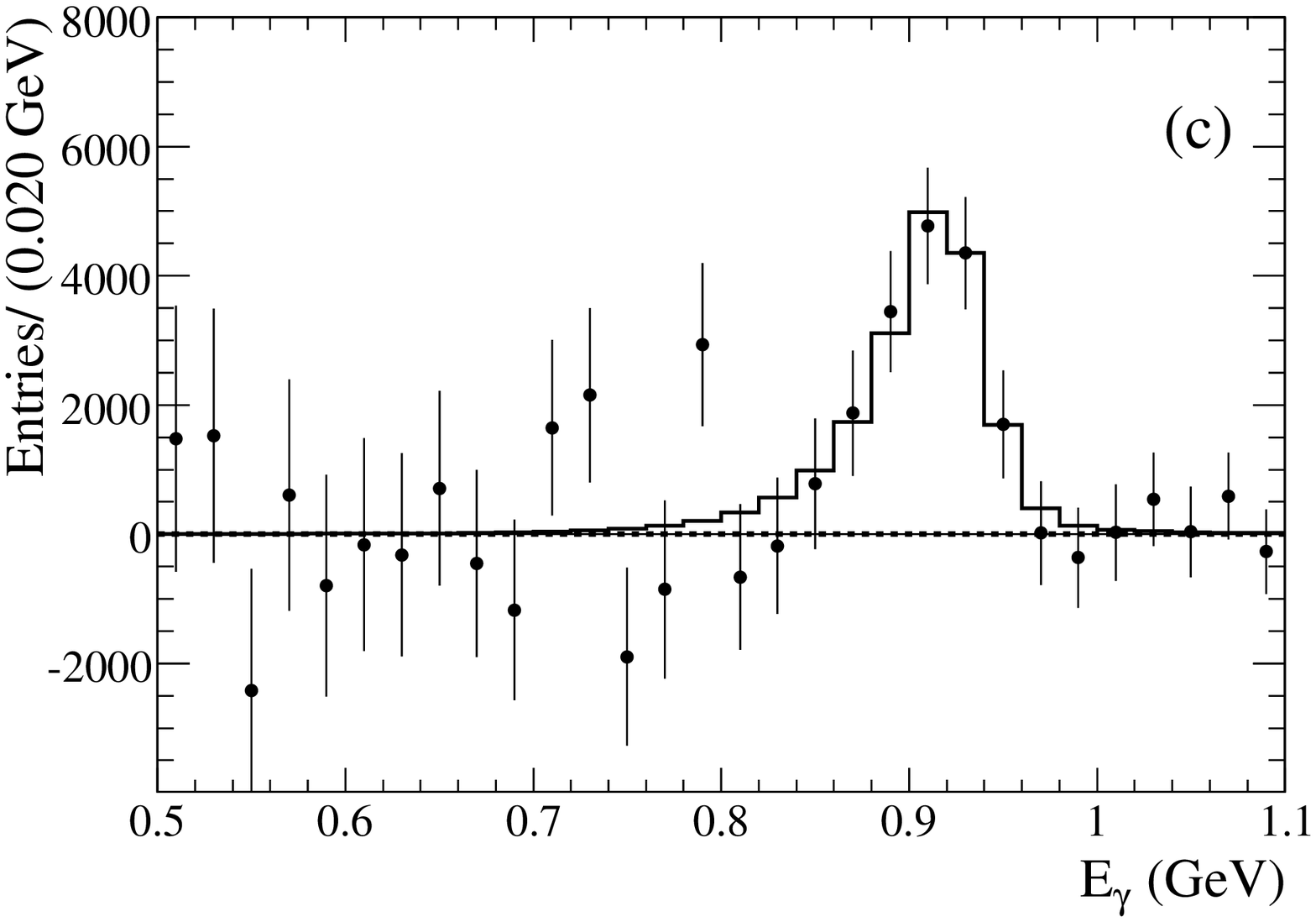}
\caption{(a) Inclusive $E_\gamma$ distribution with fit result. Only the 
$\chi_{bJ}(2P)\rightarrow \gamma \, \Upsilon(1S)$ is visible. (b) Non-peaking background 
subtracted plot, with PDFs for $\chi_{bJ}(2P)\rightarrow \gamma \, \Upsilon(1S)$ peak (solid), 
ISR peak (dot), $\eta_b$ signal (dash) and the sum of all three (solid). (c) All-background 
subtracted distribution.} \label{figfit}
\end{figure*}

The final fit to the $E_\gamma$ distribution was performed with the PDFs described above. The free 
paramaters were the $\chi_{bJ}(2P)\rightarrow \gamma \, \Upsilon(1S)$ process yield, the non-peaking 
background parameters and the signal yield. Figure \ref{figfit}(a) shows the $E_\gamma$ distribution 
and the fit result. In addition to the non-peaking background, only the 
$\chi_{bJ}(2P)\rightarrow \gamma \, \Upsilon(1S)$ broad peak is visible. Figure \ref{figfit}(b) shows 
the non-peaking background subtracted plot  in the signal region. The 
$\chi_{bJ}(2P)\rightarrow \gamma \, \Upsilon(1S)$, $\gamma_{ISR} \Upsilon(1S)$, and signal peaks 
are clearly visible. Figure \ref{figfit}(c) shows the background subtracted distribution overlaid 
with the fit result for the $\eta_b$ PDF. 

The fitted $\eta_b$ signal yield is $19200 \pm 2000 \pm 2100$ events, where the first error is 
statistical and the second systematic. The systematic error has been obtained from varying the 
$\eta_b$ width (from 5 to 20 MeV), the $\gamma_{ISR} \Upsilon(1S)$ yield within $\pm$1 $\sigma$ of 
the nominal value, and the PDF parameters within $\pm$1 $\sigma$.

The $\eta_b$ signal significance is estimated using the ratio  log$ (L_{\rm max} / L_0)$,
where $L_{\rm max}$ and $L_0$ are the likelihood values obtained from the nominal fit and from
a fit with the $\eta_b$ PDF removed, respectively. The significance the signal has been 
conservativaly estimated with the following method: a fit to the data has been performed with 
all parameters entering the systematic errors moved by 1 standard deviation in the direction of 
smallest significance. This method yields to a 10 standard deviations significance.

\section{RESULTS}

The fitted $\eta_b$ signal position is $917.4 ^{+2.1}_{-2.8}$ MeV. A photon energy calibration 
shift of $3.8 \pm 2.0 $ MeV is then applied. It is obtained by comparing the fitted position of 
the $\chi_{bJ}(2P)\rightarrow \gamma \, \Upsilon(1S)$ peak to the PDG value. Applying 
the energy calibration shift, we obtain for the peak position of the $\eta_b$ signal: 
$E_\gamma = 921.2 ^{+2.1}_{-2.8} \pm 2.4$ MeV. 

This yields to the $\eta_b$ mass: $M(\eta_b) = 9388.9 ^{+3.1}_{-2.3} \pm 2.7$ MeV/$c^2$. 
Using the PDG value of $9460.3 \pm 0.3$ MeV/$c^2$ for the $\Upsilon(1S)$ mass,
we determine the $\Upsilon(1S)$-$\eta_b$ mass splitting to be $71.4 ^{+2.3}_{-3.1} \pm 2.7$ MeV/$c^2$.

Using the signal reconstruction efficieny and the number of $\Upsilon(3S)$ events, we estimate 
the $\Upsilon(3S) \to \gamma \, \eta_b$ branching fraction to be $(4.8 \pm 0.5 \pm 1.2) \times 10^{-4}$, 
where the first uncertainty is statistical and the second systematic. The main systematic uncertainty 
is from the efficiency. We have compared the reconstruction efficiency of the 
$\chi_{bJ}(2P)\rightarrow \gamma \, \Upsilon(1S)$ peak between data and MC, giving a $18\%$ error.

\section{CONCLUSION}

In conlusion, we have made the first observation of the bottomonium ground state, 
the $\eta_b$. The new state has been observed in the radiative decay of the $\Upsilon(3S)$. 
The $\eta_b$ is the most likely interpretation of the signal, although other hypothesis 
are not excluded. 
The mass of the $\eta_b$ is $9388.9 ^{+3.1}_{-2.3} \pm 2.7$ MeV/$c^2$, which
corresponds to a mass splitting between the $\Upsilon(1S)$ and the $\eta_b$ of
$71.4 ^{+2.3}_{-3.1} \pm 2.7$ MeV/$c^2$. The estimated branching fraction of the decay
$\Upsilon(3S) \to \gamma \, \eta_b$ is found to be $(4.8 \pm 0.5 \pm 1.2) \times 10^{-4}$.

\begin{acknowledgments}
The author is representing the BaBar Collaboration. We are grateful for the excellent 
luminosity and machine conditions provided by our PEP-II colleagues, and for the substantial 
dedicated effort from the computing organizations that support BaBar. We thank Bob McElrath 
and Michael Peskin for helpful discussions. The collaborating institutions wish to thank
SLAC for its support and kind hospitality. 
This work is supported by DOE and NSF (USA), NSERC (Canada), CEA and CNRS-IN2P3 (France), 
BMBF and DFG (Germany),
INFN (Italy),
FOM (The Netherlands),
NFR (Norway),
MES (Russia),
MEC (Spain), and
STFC (United Kingdom).
Individuals have received support from the
Marie Curie EIF (European Union) and
the A.~P.~Sloan Foundation.

\end{acknowledgments}

\end{document}